\documentstyle[epsf]{article}

\newcommand{\bee}{\begin{equation}}
\newcommand{\ee}{\end{equation}}
\newcommand{\beea}{\begin{eqnarray}}
\newcommand{\eea}{\end{eqnarray}}

\begin{document}
\thispagestyle{empty}
\parskip=12pt
\raggedbottom

\def\mytoday#1{{ } \ifcase\month \or
 January\or February\or March\or April\or May\or June\or
 July\or August\or September\or October\or November\or December\fi
 \space \number\year}
\noindent
\hspace*{9cm} COLO-HEP-396\\
\vspace*{1cm}
\begin{center}
{\LARGE  Structure of the QCD Vacuum As Seen By Lattice Simulations}

\vspace{0.5cm}

T. DeGrand, Anna Hasenfratz,  Tam\'as Kov\'acs\\
Physics Department, 
        University of Colorado, \\ 
        Boulder, CO 80309 USA

\begin{abstract}
This talk is a review of our studies of instantons and their properties
as seen in our lattice simulations of $SU(2)$ gauge theory.  
We have measured the topological susceptibility and the size distribution
of instantons in the QCD vacuum.
 We have also investigated the properties
of quarks moving in instanton background field configurations, where the
sizes and locations of the instantons are taken from simulations
of the full gauge theory.
By themselves, these multi-instanton configurations do not confine quarks, but
they induce chiral symmetry breaking.
\end{abstract}

Talk presented by T. DeGrand at the 1997 Yukawa International Seminar
``Nonperturbative QCD--Structure of the QCD Vacuum''

\end{center}
\eject

\section{Introduction}

What features of the QCD vacuum are responsible for confinement or
for the generation of the observed structure of hadron spectroscopy?
  This question might,
in principle, be answered by lattice simulations of non-Abelian gauge
theories.
We have been studying the properties of instantons from $SU(2)$ lattice
simulations.
This talk is a survey of our work  as presented in
 Refs.~\cite{INSTANTON1,INSTANTON2,SU2_DENS,INSTAPE,INSTPBP}.

The study of the QCD vacuum using lattice Monte Carlo is complicated
by two problems.  The first one is that  
the dominant features of the QCD vacuum as seen
in lattice simulations are  short distance fluctuations (as they would be
for any quantum field theory).  They are basically uninteresting noise. 
 The solution to this problem is to invent operators which are 
insensitive to the short distance behavior of the field variables.
This brings the second problem: The separation of vacuum structure into
short distance and long distance parts is ambiguous, and what one 
sees can depend strongly on the operators one uses.
All direct smoothing transformations\cite{COOL}
 distort the original lattice
configuration.  This makes the extraction of (continuum) short  to medium
distance physics,
like observations of topological objects, very delicate.
If any space-time symmetric smoothing transformation
is repeated
 enough times, all the vacuum structure in any finite
volume, including the simulation volume, will be washed away.
Thus, it does not make sense to extrapolate one's results to
the limit of a very large number   of smoothing steps.
The only measurements which are physically meaningful are measurements
which are extrapolated back to  the original lattice, that is, back to
zero smoothing steps.  This requires careful monitoring of observables
over the whole history of smoothing transformations.

 In QCD instantons may be responsible for
breaking axial symmetry and resolving the $U(1)$ problem. \cite{U1}
The relevant observable is the
topological susceptibility $\chi_t$, defined as the infinite volume
 limit of
\bee
\chi_t = \langle \int d^4 x Q(x)Q(0) \rangle = {{\langle Q^2 \rangle}\over V}
\label{EQ1}
\ee
where $Q$ is the topological charge and $V$ the space time volume.
In QCD $\chi_t$ is a dimension-4 object with no weak coupling expansion,
and a calculation of $\chi_t$ in physical units in the continuum
requires nonperturbative techniques.
In the large-$N_c$ limit the mass of the $\eta'$ is (probably!)
related to the
topological susceptibility through
the Witten-Veneziano formula\cite{WV}
\bee
{f_\pi^2 \over {2 N_f}}(m^2_{\eta'} + m^2_{\eta} - 2m^2_K )=  \chi_t.
\label{WZF}
\ee
The left hand side of this equation is equal to (180 {\rm MeV})${}^4$
in the real world.

Based on phenomenological models, it has been argued that instantons
are largely responsible for chiral symmetry breaking and
the low energy hadron and glueball spectrum.
\cite{Diakanov,Shuryak_long} Instanton liquid
models attempt to reproduce the topological content of the QCD
vacuum and conclude that hadronic correlators in the instanton liquid
show all the important properties of the corresponding full
QCD correlators. These models appear to capture the essence of
the QCD vacuum, but their derivations involve a number of uncontrolled
approximations and phenomenological parameters.

 Instanton physics on the lattice is as full of controversy as continuum
instanton physics.
There are presently three different ways of measuring a topological charge.
The ``geometric'' definition\cite{GEO_O3,GEOMETRIC} reconstructs 
a fiber bundle from the
lattice gauge field and identifies the second Chern number of this bundle with
the topological charge.  It will always give an integer, 
but if the configuration is sufficiently rough, it can fail catastrophically.
``Algebraic'' definitions\cite{ALGEBRA}
introduce some lattice discretization of $Q$, as a sum of closed paths
of loops.
The worst aspect of the algebraic
 definitions is that the topological charge can mix with quantum fluctuations.
Finally, one can define $Q$ through fermionic operators. (For a recent example,
see Ref.~\cite{Narayanan}.)
All these definitions have a cutoff: they cannot see
 instanton (like) configurations when the instanton radius is too small
(typically $\rho/a \simeq 1-2$). It is usually not possible to specify
this cutoff precisely, and it can contaminate lattice measurements.

\section{Finding Instantons}

We have explored three different methods for extracting information about
topology from lattice simulations. Each has its own strengths and
weaknesses.

\subsection{Inverse Blocking}
``Inverse blocking'' does not distort the original configuration.
This technique has been introduced by Hasenfratz\cite{HASYK} in his lectures.
Imagine beginning with a set of lattice variables $\{V\}$
on a lattice whose spacing is $a$ and lattice size is $L$.
The lattice action is a fixed-point (FP) action \cite{FP} $S^{FP}(U)$.
The  inverse blocking transformation
  constructs a set of fine lattice variables
$\{U\}$ occupying a lattice of lattice spacing $a/2$ and
lattice size $2L$, by solving the
steepest-descent equation
\bee
S^{FP}(V)=\min_{ \{U\} } \left( S^{FP}(U) +\kappa T(U,V)\right),  \label{STEEP}
\ee
where  $\kappa T(U,V)$ is the blocking kernel.
Inverse blocking identifies
the smoothest among the configurations that block back to the original configuration.  Since for fixed point actions, topology is unchanged
by inverse blocking, the strategy is to take a rough configuration (generated
by Monte Carlo using a FP action), inverse block it, and measure the 
charge on the fine lattice, where the instantons are twice as big
and charge measurement is more reliable.
We did this\cite{INSTANTON2} and found
 $\chi_t$ = (235(10) MeV)${}^4$  for $SU(2)$.
This created a low-level 
controversy, since our number was larger than
other measurements.\cite{Narayanan,Forcrand,Delia}

Inverse blocking is reliable, but it is very expensive.
In four dimensions, one can only perform it once (to go from lattice spacing $a$
to $a/2$).  The fine configurations are still too rough to 
identify individual instantons.

\subsection{Cycling}
Our smoothing mechanism for seeing instantons is called ``cycling.''
\cite{SU2_DENS}
One first performs an inverse blocking from a coarse lattice to a set of
fine lattice variables by solving Eqn. \ref{STEEP}.
Now the original lattice occupies one of the 16 sublattices of
the fine lattice. Next,
 we perform a blocking transformation to a set of coarse variables
 $\{W\}$ based on one of the other sublattices. The delicate coherence
among the fine variables
is broken and the new coarse variables
are strongly ordered on the
shortest distance scale (as measured, for example, by the expectation value
of the plaquette) while retaining all long distance physics (because they
are generated by a RG blocking transformation).
This is the second part of the cycling transformation
$V_\mu(x)\rightarrow U_\mu(x) \rightarrow W_\mu(x)$.
Cycling steps can be iterated, and a few cycling steps can reduce
the plaquette to within 0.001 of its free-field value.

 Individual instantons can
be seen after a few cycling steps.  Their sizes drift with smoothing
(see the next section for pictures) but the drift is small enough that
we can extrapolate their properties back to zero cycling steps.
We found $\chi_t^{1/4}=230(10)$ MeV  for $SU(2)$,
 an instanton density of about
two per fm${}^4$, and a mean instanton radius of about 0.2 fm.
Again, the susceptibility is higher than others', and now the mean
instanton size is smaller than other measurements, although it agrees
with the predictions of the interacting instanton liquid model.
\cite{Diakanov,Shuryak_long}

Cycling is still quite expensive because of the inverse blocking step.
This makes it hard to push to small lattice spacing and to test scaling.

\subsection{RG Mapping}
Cycling produces a sequence of coarse ($\{V\}$) and fine ($\{U\}$)
lattices $ \{V\} \to \{U_1\} \to \{V_1\} \to \{U_2\} \to \{V_2\} \to \dots$.
RG mapping is a technique for eliminating the inverse blocking step
 and generating
a sequence of coarse 
lattices $ \{V\} \to \{W_1\} \to \{W_2\} \to \dots$
where $\{W_n\}$ is an approximation to $\{V_n\}$.
 The idea is that,
while formally the inverse blocking is  non-local, for local FP actions the
dependence of the fine
links on the original coarse links dies away exponentially
with their separation \cite{FP} and the mapping
$\{V\} \to \{U_1(V)\} \to \{V_1(V)\}$ 
can be considered local. The $\{W_n\}$ lattice
can be constructed from the original  coarse $\{V\}$ lattice as a sum
of loops of $\{V\}$'s which are designed to reproduce a cycled
$\{V_n\}$.
We discovered that this could be done by
APE-smearing:  \cite{APEBlock} from a set $\{V\}$ construct a new set of
links $\{X\}$ by
\beea
X_\mu(x) = (1-c)V_\mu(x) & +  & c/6 \sum_{\nu \ne \mu}
(V_\nu(x)V_\mu(x+\hat \nu)V_\mu(x+\hat \nu)^\dagger
\nonumber  \\
& + & V_\nu(x- \hat \nu)^\dagger
 V_\mu(x- \hat \nu)V_\mu(x - \hat \nu +\hat \mu) ),
\label{APE}
\eea
with  $X_\mu(x)$  projected back onto $SU(2)$ to generate  $W_\mu(x)$.
We found that a series of steps with $c=0.45$ was a good choice to mock up
cycling.

RG-mapping is very cheap, but the price is that everything 
which is measured must be monitored carefully, and  extrapolated (if necessary)
back to zero mapping steps.
 
The problem is that any approximation to the fixed point charge is 
expected to distort the charge density profile. We have to correct
this distortion.
This can be done by monitoring how
the charge density--measured by the FP charge--changes
in the course of smoothing and extrapolating this back to zero smoothing
steps. Of course we cannot directly measure the FP charge since this
would involve several inverse blocking steps. However after a few smoothing
steps the configurations become smooth enough so that our improved charge
operator is very close to the exact FP charge and data taken after further
smoothing can be used for extrapolating back. 
 
\begin{figure}
\epsfxsize = 8 cm
\centerline{\epsfbox{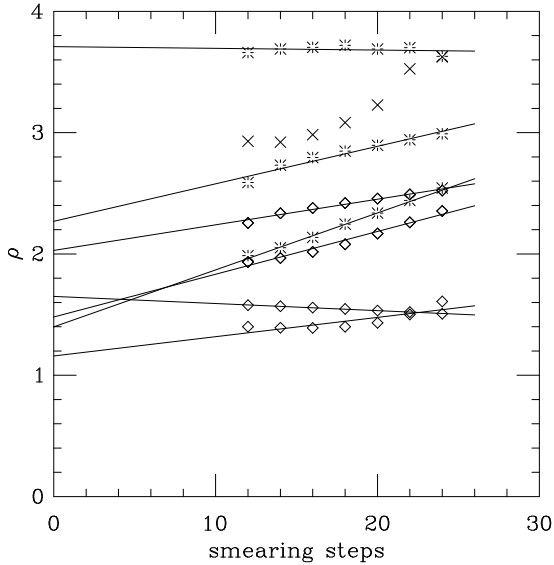}
}
\caption{ Radius versus APE-smearing steps of instantons (diamonds) and
anti-instantons (bursts and crosses) on a $16^4$ $\beta=2.5$ configuration.}
\label{fig:rho_vs_APE}
\end{figure}

As an example, we consider a $16^4$ configuration generated with
Wilson action $\beta=2.5$.
Fig. \ref{fig:rho_vs_APE} shows the size of the 8 ``stable'' objects as a function of the
APE-smearing steps. From the 4 instantons (diamonds) 3 increase in size
while one
decreases, but  all vary linearly with smearing steps.
The slope of the linear change for all of them is small. Three of the 
anti-instantons (bursts) behave similarly, though one has a slightly larger slope. The fourth anti-instanton
(crosses) starts to grow rapidly after 18 smearing steps and will
disappear after a few more steps. This object is likely to
be a vacuum fluctuation, not an instanton.

In the early stages of cycling (1-2 steps) we see many lumps of charge.
Most of them quickly grow or shrink away. The locations of what we call  true
topological objects are stable over many smearing steps
and their size changes slowly. To identify
them on the lattice one has to track them over several smearing steps and monitor
their behavior.

Instantons present in QCD simulations differ from
 hand crafted instantons in trivial background configurations
in that the former usually grow  while the latter objects
always  shrink under APE
smearing.
 
Neither cycling nor RG-mapping affect the string tension, but the short 
distance part of the potential is distorted. 

 
One of our goals was to compare results obtained with the cycling/RG
mapping method with results published using other algorithms
\cite{MS,Forcrand,Delia}, so we used the Wilson action in conjunction
with RG-mapping.

We observe, as expected, a
small systematic decrease in the susceptibility as we increase the
number of smearing steps.
At large $\beta$ the change is small and statistically insignificant.
Only at $\beta=2.4$, where the configurations are the roughest,
do the 12 and 24 smearing steps results differ by about  a standard
deviation.

The susceptibility increases by 10\% from $\beta=2.4$ to $\beta=2.5$
but stabilizes after that at the value $\chi_t^{1/4}=220(6)$ MeV.
We interpret the change between  the extrapolated $\beta=2.4$ result
and the larger $\beta$ results as due to the absence of
small  instantons  at $\beta=2.4$ because of the larger lattice spacing. This
interpretation will be supported by the instanton size distribution result
discussed below.

\begin{figure}
\epsfxsize = 8 cm
\centerline{\epsfbox{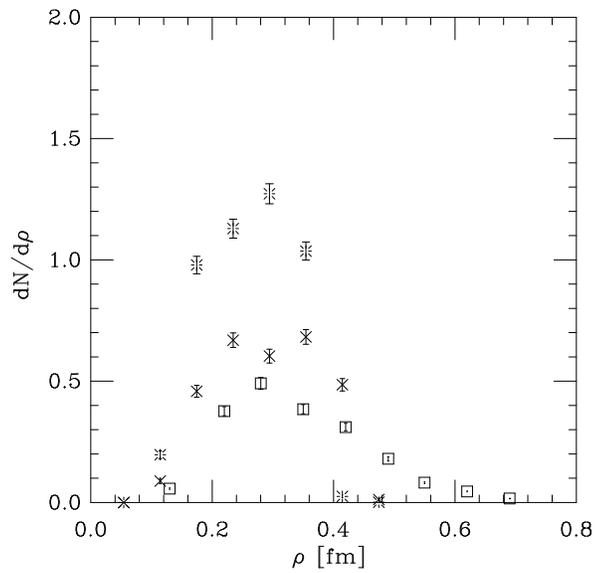}}
\caption{The size distribution on 12 (bursts) and 24 (crosses) times APE smeared lattices at
$\beta=2.5$. The square symbols are the results of Ref. 11 rescaled appropriately.}
\label{fig:d_rho_direct}
\end{figure}

We have identified individual instantons after every 2 APE-smearing steps between
12 and 24
steps. Since these configurations are still rough, many of the objects identified
as
instantons are in fact vacuum fluctuations and disappear
 after more smoothing steps. 
 Figure \ref{fig:d_rho_direct} shows the observed
instanton
size distribution on the 12 and 24
times smeared lattices at $\beta=2.5$. 
For comparison we also plot the result of Ref.~\cite{MS}
 which corresponds, in our normalization,
 to about 100 times smeared lattices. It is
obvious from the figure that the total number of identified objects decreases
 as we increase
the smoothing.
The density of identified objects is 4.6 per  fm$^4$ after 12
smearing steps, 3.0 per fm$^4$ after 24
smearing steps, and about 2 per fm$^4$ in Ref.~\cite{MS}. These density values are considerably larger than the expected value of about 1 per fm$^4$.
The maximum of all 3 distributions is around $\bar \rho \approx 0.3$ fm, but
that does not mean that on the original lattice  $\bar \rho \approx
0.3$ fm since instantons usually grow under smearing. This
growth can also  be observed from the increasing tail of the
distribution, especially after the many blocking steps of Ref.~\cite{MS}.

\begin{figure}
\epsfxsize = 8 cm
\centerline{\epsfbox{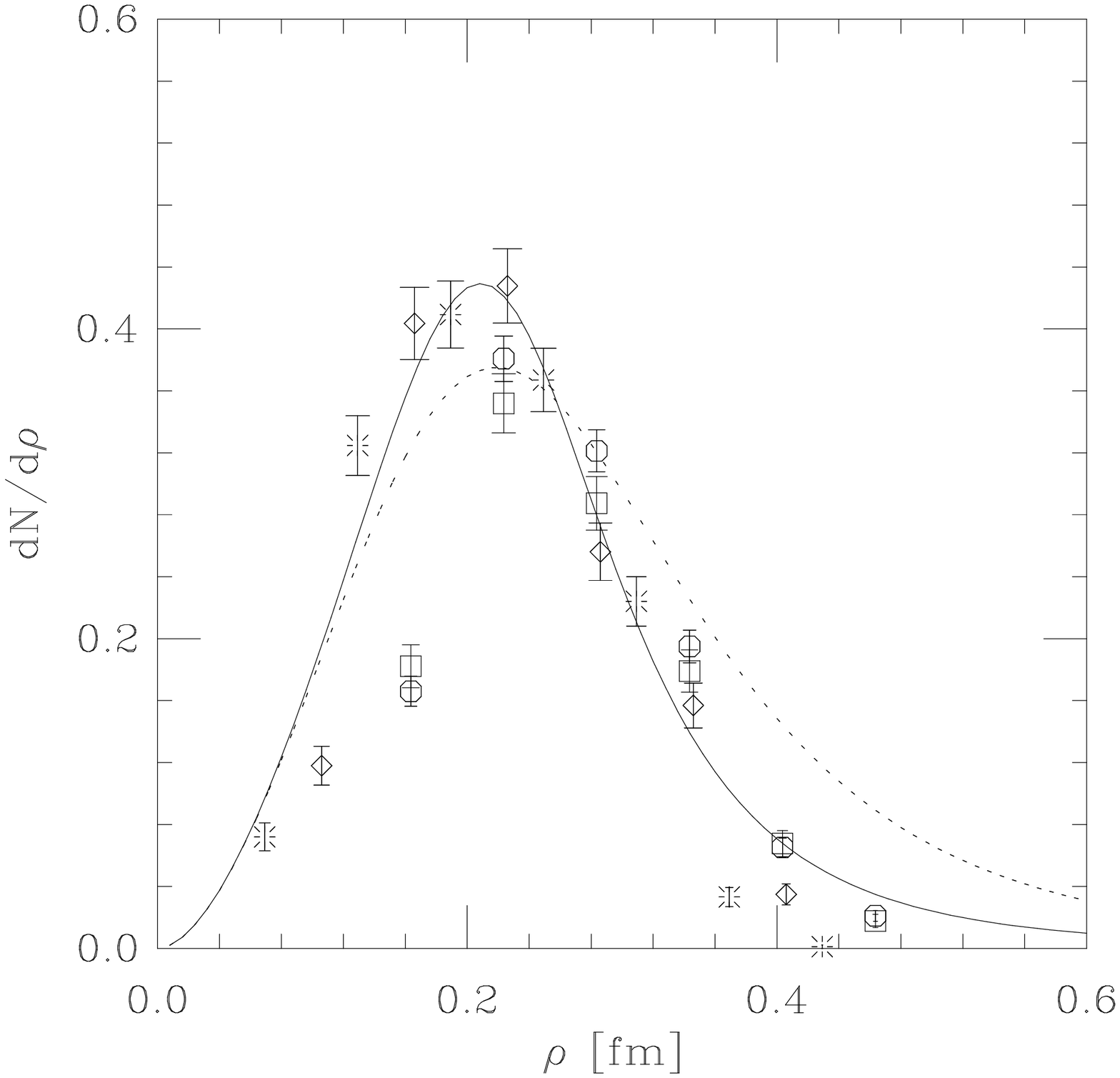}}
\caption{The size distribution of instantons. Octagons correspond to $\beta=2.4$ $12^4$,
squares  to $\beta=2.4$ $20^4$, diamonds
to $\beta=2.5$ and bursts to $\beta=2.6$. The first bin of each distribution is
contaminated by the cut-off.
The solid
curve is a two parameter fit to the data points according to the formula in Ref. 23.
The dashed curve is a similar fit from Ref. 23 which describes the instanton liquid model quite
closely. }
\label{fig:d_rho_wil}
\end{figure}

Our final result for the instanton size
distribution
(after extrapolation back to zero smoothing steps)
 is shown in Figure \ref{fig:d_rho_wil}, where
we overlay the data obtained at $\beta=2.4$ (octagons on $12^4$, squares on $20^4$ lattices),
$\beta=2.5$ (diamonds), and
$\beta=2.6$ (bursts).
Since the smoothing method cannot identify instantons with
$\rho\leq 1.5a$, we chose the bins such that the second
 bin for each distribution starts at
$\rho=1.6a$. That means that we expect the second bin of each distribution to be universal. The first bins on the
other hand contain only some of the small instantons and their value is not expected to be
universal.

The four distributions form a universal curve
indicating scaling. The $\beta=2.4$ curves cover
only the $\rho>0.2$ fm region,
and small instantons are obviously missing. The agreement between the $12^4$ and $20^4$
configurations at $\beta=2.4$ indicate that a linear size of about 1.4 fm is sufficient to
observe all the topological objects. The $\beta=2.5$ and
2.6 distributions have most of the physically relevant instantons,
 supporting the scaling
behavior observed for the topological susceptibility.
 
The instanton liquid model predicts a very similar picture to ours.
In Ref.~\cite{Shur.distrib}
 Shuryak predicted an instanton size distribution that peaked
around $\rho=0.2$ fm. The curve in Fig. \ref{fig:d_rho_wil} 
is a fit using the two loop perturbative instanton distribution formula with a
``regularized'' log
\beea
S_I={8\pi^2 \over g^2(\rho)} = b_0L + b_1 \log L \\
L={1 \over p}\log[(\rho\Lambda_{inst})^{-p}+C^{p}]
\eea
where $b_0$ and $b_1$ are related to the first two coefficients of the
perturbative $\beta$ function and $p$
and $C$ are arbitrary parameters.
The solid
curve is a fit to our data while the dashed curve is
the fit given in Ref.~\cite{Shur.distrib} which
 describes the instanton liquid model quite
closely.  The
difference between the two fits is significant at large $\rho$ values only.
We do not know if changing the parameters of the interacting instanton liquid model slightly would change
the predictions of that model improving the agreement with the Monte Carlo data.
 
 To make a long story short, we believe that there are two reasons why
our instanton numbers are different from those of
 Refs.~\cite{MS,Forcrand,Delia}. First, we are sensitive to instantons of 
smaller radius. As we cut larger and larger instantons from our sample,
we see $\chi_t$ fall.  Second (and most important) we extrapolate our
measurements back to zero smoothing steps. Instantons generally grow under
smoothing, so our extrapolated sizes are smaller.

\section{What Do Instantons Do?}

To test what role instantons play in the QCD vacuum, we took a set of $SU(2)$ configurations  with lattice spacing $a \simeq 0.14$ fm
 and smoothed them enough to identify the
instantons. These smoothed configurations have essentially the same string
tension as the original configurations, even though about seventy per
cent of their vacuum 
action is carried by the instantons.  We then
identified the sizes and locations of the instantons in the configurations
and built multi-instanton configurations from them. We built both
``parallel'' and randomly oriented instanton configurations.

Notice that the instanton locations are not random, and the sizes are not
taken from a model distribution--they come from the (lattice) QCD vacuum.

\begin{figure}
\epsfxsize = 8 cm
\centerline{\epsfbox{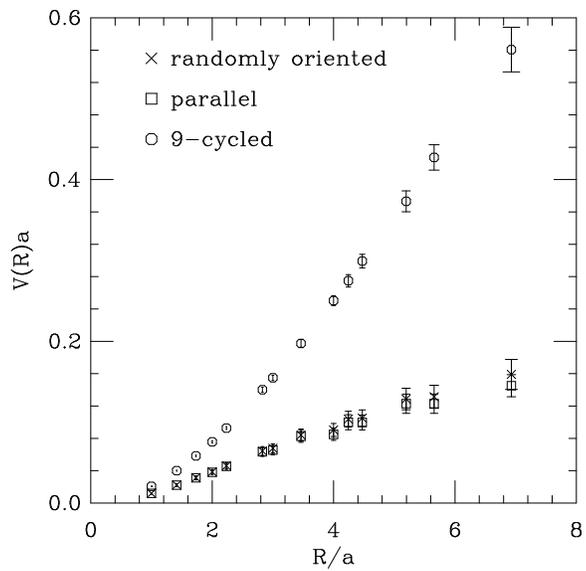}}
\caption{The heavy-quark potential measured on the 9 times cycled
real configurations (octogons), the randomly (crosses) and the parallel
(squares) oriented instantons.}
\label{fig:pot_inst_all}
\end{figure}

In Fig.\ \ref{fig:pot_inst_all} the heavy-quark potentials obtained from the 
three ensembles are compared. We can conclude that neither the parallel
nor the randomly oriented instantons confine. 
 It seems that instantons by themselves are
not responsible for confinement.

Next, we computed hadron spectroscopy in these background configurations.
One could think of spectroscopy on the smoothed configurations as spectroscopy
computed with a new kind of quark action, one which is insensitive to the
short distance behavior of the gauge field.
This is similar in spirit to the use of ``fat links'' by
the MILC collaboration \cite{MILCIMPR} 
and by  J.-F. Laga\"e and D.~K. Sinclair \cite{SINCLAIR},
or of the approximate FP actions developed by one of us.\cite{TOM}

Staggered fermions provide the most interesting results.
It appears that cycling does not affect  the $\rho$ mass vs the
Goldstone pion.  However, on the smoothed lattices
the non-Goldstone partner of the pion is degenerate with the pion
within observational uncertainty.
On the original lattices the two mesons have a mass ratio of about 1.4.

\begin{figure}
\epsfxsize = 8 cm
\centerline{\epsfbox{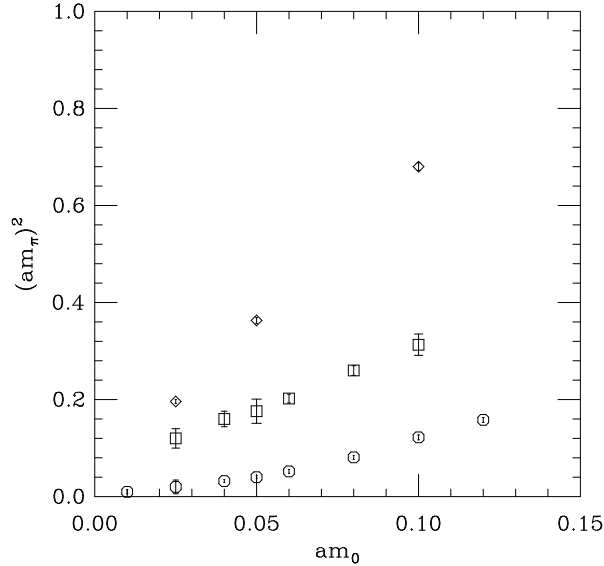}}
\caption{$(am_\pi)^2$ vs. $m_0$ the bare quark mass
for the staggered action,
with raw gauge links (diamonds) and 9-cycled links (squares).
The lightest mass in the pseudoscalar channel with instanton background
configurations is shown by octagons.}
\label{fig:pi20k}
\end{figure}

The purpose of this exercise is not to test spectroscopy in detail.
The important point is that the
 action of the many-times-cycled configurations is dominated by
instantons, and yet   the asymptotic heavy quark
potential and hadron spectroscopy 
are basically unchanged from what we saw on untouched configurations.

But these results do not answer the question:
 Are instantons responsible for long distance physics,
or are the structures responsible for long distance physics some other
objects carrying low action, which have been preserved by the smoothing transformation?

We  computed spectroscopy on the two types of
multi-instanton ensembles using both  staggered and Wilson fermions.
The dominant feature of both spectroscopy calculation is that
the quarks are deconfined. This is seen most easily in
the staggered fermion spectroscopy. 

\begin{figure}
\epsfxsize = 8 cm
\centerline{\epsfbox{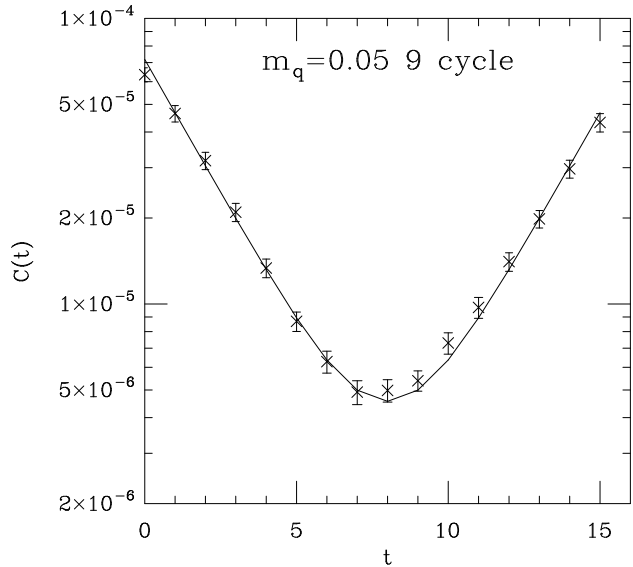}}
\caption{The pseudoscalar propagator from 9-cycled configurations, with
staggered fermions of bare mass $am_0=0.05$.}
\label{fig:pseudo905}
\end{figure}

\begin{figure}
\epsfxsize = 8 cm
\centerline{\epsfbox{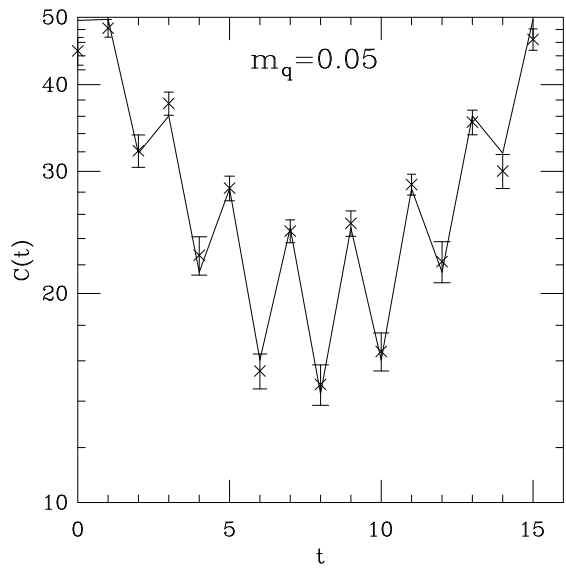}}
\caption{The pseudoscalar propagator in (randomly rotated)
 instanton background
 configurations, with
staggered fermions of bare mass $am_0=0.05$.
The curve is a fit to a single propagating particle plus the
$q \bar q$ branch cut.}
\label{fig:pseudo05}
\end{figure}

Consider the pseudoscalar propagator of the 9-cycled configurations,
shown for one quark mass in Fig. \ref{fig:pseudo905}. It looks like 
any generic lattice pseudoscalar, a more-or-less pure hyperbolic
cosine with no oscillations.
The staggered fermion pseudoscalar propagators on instanton background fields
are quite different: they show the characteristic
sawtooth pattern of free antiperiodic  fermions. (See Fig.
 \ref{fig:pseudo05}).

\begin{figure}
\epsfxsize = 8 cm
\centerline{\epsfbox{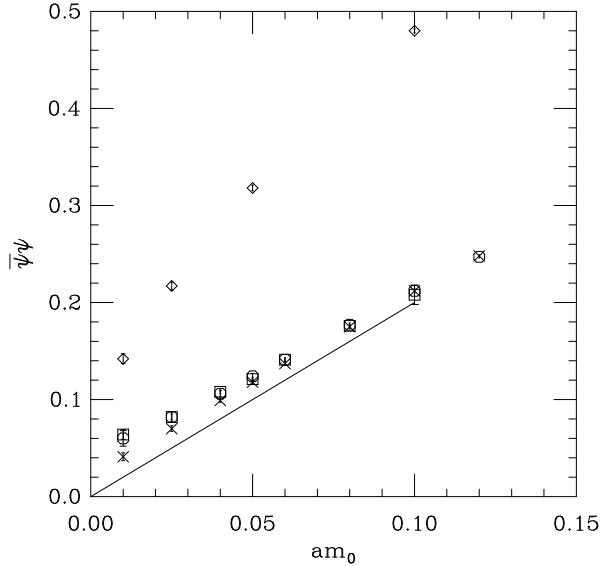}}
\caption{$\langle \bar \psi \psi \rangle$ from raw configurations (diamonds),
 9-cycled configurations (squares), and  instanton-background configurations
which are parallel (crosses) and randomly oriented (octagons),
vs. bare quark mass $am_0$. The line shows the free-field value, $2m_0$.}
\label{fig:pbp}
\end{figure}

 Fig.  \ref{fig:pbp} shows the lattice
$\langle \bar \psi \psi \rangle$ for staggered fermions on the 9-cycled
and in instanton background
configurations. $\langle \bar \psi \psi \rangle$ in the instanton background 
tracks the value of $\langle \bar \psi \psi \rangle$
measured on the 9-cycled configurations quite closely, down to small
quark mass.
It appears that the instantons,
present in equilibrium gauge field configurations of the QCD vacuum
generated
by Monte Carlo,
are breaking chiral symmetry by themselves.
This effect is  a cornerstone of
instanton-liquid models of hadron structure.

If  $\langle \bar \psi \psi \rangle$ is nonzero, one expects the spectrum
contains a would-be Goldstone boson (the pion) in addition to
massive quarks and (possibly) other resonances.
To test this hypothesis,  we fit the pseudoscalar correlator to two terms:
a pure hyperbolic cosine (a pole in the frequency plane), plus
a $q \bar q$ branch cut, with the quark mass as a parameter. We used
the
analytic expression for the branch cut
(expressed as a momentum and frequency mode sum),
 with the appropriate boundary conditions
and source used in the simulations.
In the random instanton background,
we clearly see a light mass which decreases towards zero
as the quark mass vanishes. (See Fig.~\ref{fig:pi20k}.)

The quark mass in the randomly oriented instanton
background is also   determined by the fit.
It  is close to the bare mass and is
 less than half the rho mass--and less than half the
pion mass-- (as measured on the smoothed lattices).

Are there other bound states? Only for the randomly rotated 
instanton configurations could we see a convincing pseudoscalar,
and so we restricted our analysis to that data set.  We again attempted
to fit to a single resonance plus a $q\bar q$ continuum.
Of course, this assumption is questionable. 
The fits are not of high quality and our results should not be taken
too seriously:  In the ``SC'' channel (saturated by the $\pi_2$ and scalar
mesons in the confined phase) we saw a light bound state whose mass
roughly tracked the mass of the (presumed) pion resonance in the
 pseudoscalar channel. (That is, there is a multiplet of Goldstone bosons.)
In the vector channels, a state with a mass 750-850 MeV appears, in addition to
the free $q\bar q$ continuum.  This state has about the same mass
as the vector meson in the confined system.
However, the dominant feature of all these channels is still the
free $q\bar q$ continuum, with fitted quark masses of 100 MeV or lower
whose energy  is always lower than the mass
of any (presumed) resonance.

The reader should be aware that the lattice spacing is large enough that it
could contaminate the coupling of quarks to instantons.

\section{Conclusions}

We have measured the instanton content of the $SU(2)$ lattice vacuum.
Instantons do not confine, but they seem to be connected to chiral symmetry
breaking.  We are presently using RG-mapping to study the properties
of instantons in $SU(3)$, both in the pure gauge theory and for full
QCD.  This spring we expect to begin computing hadron spectroscopy in smoothed
configurations and in multi-instanton background configurations.
How similar will they be?

\section*{Acknowledgements}
T.~D. would like to thank the organizers of YKIS 97 for putting together
such a diverse and interesting meeting, and for their hospitality in Kyoto.
This work was supported by the U.S. Department of Energy.

\newcommand{\PL}[3]{{Phys. Lett.} {\bf #1} {(19#2)} #3}
\newcommand{\PR}[3]{{Phys. Rev.} {\bf #1} {(19#2)}  #3}
\newcommand{\NP}[3]{{Nucl. Phys.} {\bf #1} {(19#2)} #3}
\newcommand{\PRL}[3]{{Phys. Rev. Lett.} {\bf #1} {(19#2)} #3}
\newcommand{\PREPC}[3]{{Phys. Rep.} {\bf #1} {(19#2)}  #3}
\newcommand{\ZPHYS}[3]{{Z. Phys.} {\bf #1} {(19#2)} #3}
\newcommand{\ANN}[3]{{Ann. Phys. (N.Y.)} {\bf #1} {(19#2)} #3}
\newcommand{\HELV}[3]{{Helv. Phys. Acta} {\bf #1} {(19#2)} #3}
\newcommand{\NC}[3]{{Nuovo Cim.} {\bf #1} {(19#2)} #3}
\newcommand{\CMP}[3]{{Comm. Math. Phys.} {\bf #1} {(19#2)} #3}
\newcommand{\REVMP}[3]{{Rev. Mod. Phys.} {\bf #1} {(19#2)} #3}
\newcommand{\ADD}[3]{{\hspace{.1truecm}}{\bf #1} {(19#2)} #3}
\newcommand{\PA}[3] {{Physica} {\bf #1} {(19#2)} #3}
\newcommand{\JE}[3] {{JETP} {\bf #1} {(19#2)} #3}
\newcommand{\FS}[3] {{Nucl. Phys.} {\bf #1}{[FS#2]} {(19#2)} #3}

\end{document}